\documentclass[10pt,journal,compsoc]{IEEEtran}

\usepackage{graphicx}
\usepackage{svg}
\usepackage{pdfpages}
\usepackage{wrapfig}
\usepackage{multirow}
\usepackage{multicol}
\usepackage{booktabs}
\usepackage{array}
\usepackage{threeparttable}
\usepackage[edges]{forest}
\usepackage{tikz}
\usepackage{xcolor}
\usepackage{amsmath,amsfonts}
\UseRawInputEncoding
\usepackage{url}
\usepackage{hyperref}

\ifCLASSOPTIONcompsoc
  \usepackage[nocompress]{cite}
\else
  \usepackage{cite}
\fi
\ifCLASSINFOpdf
\else
\fi
\begin{document}
%
\title{Task Scheduling in Geo-Distributed\\ Computing: A Survey}
%
%
%
%
\date{\today}

\author{
        Yujian~Wu,
        Shanjiang~Tang,
        Ce~Yu,
        Bin~Yang,
        Chao~Sun,
        Jian~Xiao,
        Hutong~Wu
\thanks{Y.J. Wu, S.J. Tang, C. Yu, B. Yang, C. Sun, J. Xiao, H.T. Wu are with the College of Intelligence and Computing, Tianjin University, Tianjin 300072, China.}%
\thanks{E-mail: \{wyuj, tashj, yuce, yangbincic, sch, xiaojian, wht\}@tju.edu.cn.}%
\thanks{Shanjiang Tang is the corresponding author.}%
\thanks{Manuscript received January 26, 2025;}}

%
%

\markboth{IEEE TRANSACTIONS ON KNOWLEDGE AND DATA ENGINEERING, VOL. XX,NO. XX, XXX. XXXX}%
{Shell \MakeLowercase{\textit{et al.}}: A survey on Geo-distributed Scheduling Optimization Techniques:}
%



\IEEEtitleabstractindextext{%
\begin{abstract}
Geo-distributed computing, a paradigm that assigns computational tasks to globally distributed nodes, has emerged as a promising approach in cloud computing, edge computing, cloud-edge computing and supercomputer computing (HPC). It enables low-latency services, ensures data locality, and handles large-scale applications. As global computing capacity and task demands increase rapidly, scheduling tasks for efficient execution in geo-distributed computing systems has become an increasingly critical research challenge. It arises from the inherent characteristics of geographic distribution, including heterogeneous network conditions, region-specific resource pricing, and varying computational capabilities across locations. Researchers have developed diverse task scheduling methods tailored to geo-distributed scenarios, aiming to achieve objectives such as performance enhancement, fairness assurance, and fault-tolerance improvement. This survey provides a comprehensive and systematic review of task scheduling techniques across four major distributed computing environments, with an in-depth analysis of these approaches based on their core scheduling objectives. Through our analysis, we identify key research challenges and outline promising directions for advancing task scheduling in geo-distributed computing.
\end{abstract}

}

\maketitle

\IEEEdisplaynontitleabstractindextext

%
\IEEEpeerreviewmaketitle

\ifCLASSOPTIONcompsoc
\IEEEraisesectionheading{\section{Introduction}\label{sec:introduction}}
\else
\section{Introduction}

\label{sec:introduction}
\fi

\begin{figure*}[ht]
    \includegraphics[width=\textwidth, page=1]{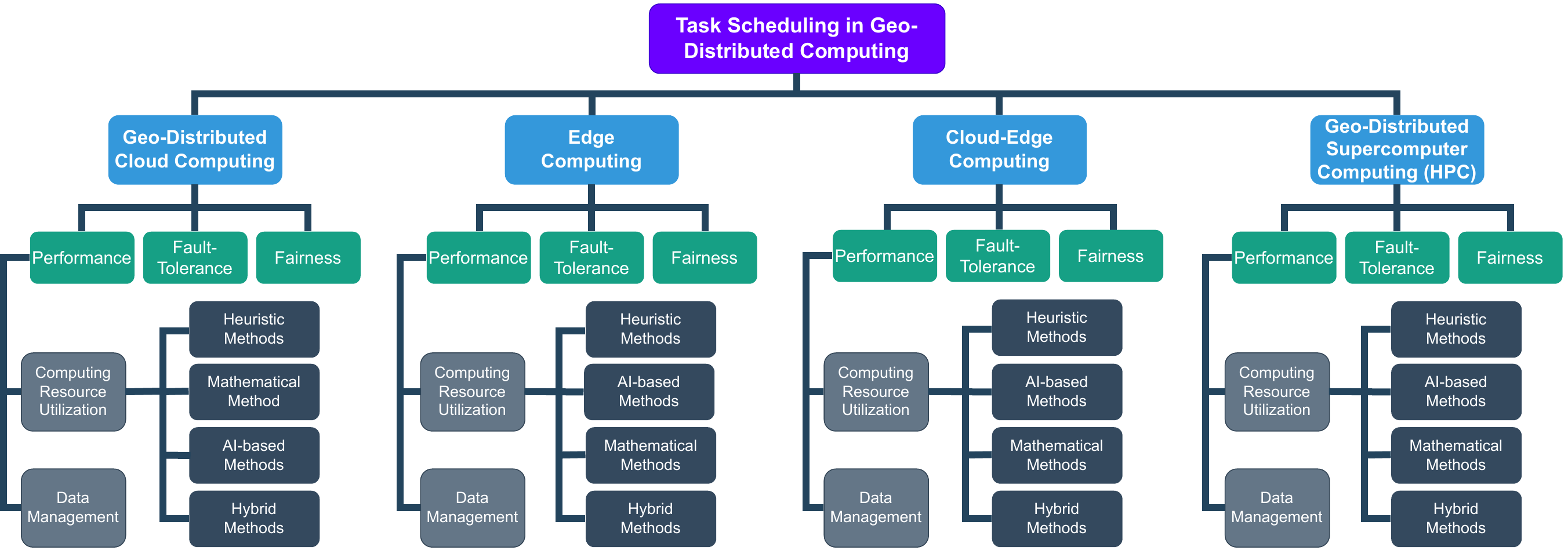}
    \caption{An overview of geo-distributed task scheduling. We categorize scheduling strategies across Geo-Distributed Cloud Computing, Cloud-Edge Computing, Edge Computing, and Geo-Distributed Supercomputer Computing (HPC). In each scheduling infrastructure, we focus on objectives including performance, fault tolerance, and fairness, with scheduling methods including heuristic, AI-based, mathematical, and hybrid techniques.}
    \label{fig:Overview_of_geo-distributed_task_scheduling}
\end{figure*}

\IEEEPARstart{I}{n} recent years, driven by the increasing distributed processing capacities and application requirements, geo-distributed computing has emerged as a new paradigm in diverse computing environments. From large-scale social networks processing billions of daily interactions to privacy-preserving federated learning systems and latency-sensitive Internet of Things (IoT) applications, modern systems inherently require computation and data processing across geographical locations. Frequently, the relevant data for these computational tasks and the computing nodes they occupy are geographically distributed.

Geo-distributed computing distributes tasks across multiple locations to enable global scalability, leverage computational capacity and provide geographical redundancy for enhanced reliability. The paradigm also minimizes user-perceived latency by processing data closer to its source, and inherently supports regional data locality requirements that many modern applications demand. However, these characteristics also introduce unique challenges in resource management and data transfer that traditional centralized scheduling systems do not encounter. Moreover, each geo-distributed computing scenario has its unique characteristics and challenges. Fully utilizing computation capacities requires more customized solutions for each environment to ensure efficient task execution.



Different geo-distributed computing environments demand distinct scheduling strategies to balance among latency, workload, and network bandwidth. Table~\ref{Tab:The_comparsion_of_different_computing_environments} illustrates these differences to better understand the unique features and requirements in each computing infrastructure. Researchers have developed numerous scheduling algorithms tailored for these geo-distributed computing systems, aiming to reduce overall makespan, minimize data transfer costs, or ensure fairness, fault-tolerance in scheduling. These approaches incorporate a range of techniques, including heuristic methods, mathematical models, and AI-based models, to enhance the execution performance of geo-distributed systems.

\begin{table}[h]
    \centering
    \caption{A COMPARISON OF DIFFERENT GEO-DISTRIBUTED COMPUTING ENVIRONMENTS}
    \label{Tab:The_comparsion_of_different_computing_environments}
    \resizebox{\columnwidth}{!}{
    \begin{tabular}{|>{\centering\arraybackslash}m{0.2\columnwidth}|>{\centering\arraybackslash}m{0.2\columnwidth}|>{\centering\arraybackslash}m{0.2\columnwidth}|>{\centering\arraybackslash}m{0.2\columnwidth}|>{\centering\arraybackslash}m{0.2\columnwidth}|}
        \toprule
        \multicolumn{1}{>{\centering\arraybackslash}m{0.2\columnwidth}}{\textbf{Feature}} & 
        \multicolumn{1}{>{\centering\arraybackslash}m{0.2\columnwidth}}{\textbf{GDCC}} & 
        \multicolumn{1}{>{\centering\arraybackslash}m{0.2\columnwidth}}{\textbf{EC}} & 
        \multicolumn{1}{>{\centering\arraybackslash}m{0.2\columnwidth}}{\textbf{CEC}} & 
        \multicolumn{1}{>{\centering\arraybackslash}m{0.2\columnwidth}}{\textbf{GDSC (HPC)}} \\
        \midrule 
        \multicolumn{1}{>{\centering\arraybackslash}m{0.2\columnwidth}}{Respond Latency} & 
        \multicolumn{1}{>{\centering\arraybackslash}m{0.2\columnwidth}}{High Latency} & 
        \multicolumn{1}{>{\centering\arraybackslash}m{0.2\columnwidth}}{Moderate Latency} & 
        \multicolumn{1}{>{\centering\arraybackslash}m{0.2\columnwidth}}{Low Latency} & 
        \multicolumn{1}{>{\centering\arraybackslash}m{0.2\columnwidth}}{Extreme Low} \\
        \midrule
        \multicolumn{1}{>{\centering\arraybackslash}m{0.2\columnwidth}}{\parbox{0.2\columnwidth}{\centering{Workload \\ Size}}} & 
        \multicolumn{1}{>{\centering\arraybackslash}m{0.2\columnwidth}}{\parbox{0.2\columnwidth}{\centering {Virtually \\ Unlimited}}} & 
        \multicolumn{1}{>{\centering\arraybackslash}m{0.2\columnwidth}}{\parbox{0.2\columnwidth}{\centering {Relatively Small}}} & 
        \multicolumn{1}{>{\centering\arraybackslash}m{0.2\columnwidth}}{\parbox{0.2\columnwidth}{\centering {Moderate Workload}}} & 
        \multicolumn{1}{>{\centering\arraybackslash}m{0.2\columnwidth}}{\parbox{0.2\columnwidth}{\centering {Exascale, \\ Heavy}}} \\
        \midrule
        \multicolumn{1}{>{\centering\arraybackslash}m{0.2\columnwidth}}{\parbox{0.2\columnwidth}{\centering {Performance}}} & 
        \multicolumn{1}{>{\centering\arraybackslash}m{0.2\columnwidth}}{\parbox{0.2\columnwidth}{\centering {High, \\ Scalable}}} & 
        \multicolumn{1}{>{\centering\arraybackslash}m{0.2\columnwidth}}{\parbox{0.2\columnwidth}{\centering {Low, \\ Limited}}} & 
        \multicolumn{1}{>{\centering\arraybackslash}m{0.2\columnwidth}}{\parbox{0.2\columnwidth}{\centering {Moderate}}} & 
        \multicolumn{1}{>{\centering\arraybackslash}m{0.2\columnwidth}}{\parbox{0.2\columnwidth}{\centering {Exceptional, \\ Scalable}}} \\
        \midrule
        \multicolumn{1}{>{\centering\arraybackslash}m{0.2\columnwidth}}{\parbox{0.2\columnwidth}{\centering {Bandwidth}}} & 
        \multicolumn{1}{>{\centering\arraybackslash}m{0.2\columnwidth}}{\parbox{0.2\columnwidth}{\centering {High \\ Demand}}} & 
        \multicolumn{1}{>{\centering\arraybackslash}m{0.2\columnwidth}}{\parbox{0.2\columnwidth}{\centering {Low \\ Demand}}} & 
        \multicolumn{1}{>{\centering\arraybackslash}m{0.2\columnwidth}}{\parbox{0.2\columnwidth}{\centering {Reduced \\ Demand}}} & 
        \multicolumn{1}{>{\centering\arraybackslash}m{0.2\columnwidth}}{\parbox{0.2\columnwidth}{\centering {Very High \\ Demand}}} \\
        \midrule
        \multicolumn{1}{>{\centering\arraybackslash}m{0.2\columnwidth}}{\parbox{0.2\columnwidth}{\centering {Task Type}}} & 
        \multicolumn{1}{>{\centering\arraybackslash}m{0.2\columnwidth}}{\parbox{0.2\columnwidth}{\centering {General-Purpose (E.g., Web Services)}}} & 
        \multicolumn{1}{>{\centering\arraybackslash}m{0.2\columnwidth}}{\parbox{0.2\columnwidth}{\centering {Real-Time, \\ Latency-Sensitive}}} & 
        \multicolumn{1}{>{\centering\arraybackslash}m{0.2\columnwidth}}{\parbox{0.2\columnwidth}{\centering {Latency-Sensitive \& Compute-Intensive}}} & 
        \multicolumn{1}{>{\centering\arraybackslash}m{0.2\columnwidth}}{\parbox{0.2\columnwidth}{\centering {Scientific Computing, Large-Scale}}} \\
        \bottomrule
    \end{tabular}
    }
    \vspace{0.5em}  
    \raggedright  
    {
    \footnotesize \\
    * GDCC: Geo-Distributed Cloud Computing.\\
    * EC: Edge Computing. CEC: Cloud-Edge Computing.\\
    * GDSC: Geo-Distributed Supercomputer Computing. 
    }
\end{table}

Despite these advancements, task scheduling in geo-distributed environment remains an active and challenging area of research. Ongoing efforts focus on developing more efficient scheduling strategies, integrating emerging computing environment like IoT, cloud-edge, and supporting new application paradigms like microservices. Furthermore, improving the usability and manageability of these systems is crucial, as it impacts the broader adoption and effectiveness of computing solutions in real-world applications. Still, with the emergence of new hardware, task scheduling in heterogeneous computing environment presents new opportunities and challenges in maximizing hardware utilization to enhance performance efficiency and system generality.

Many recent task scheduling surveys in cloud or edge computing have classified and compared scheduling strategies by algorithm types (e.g., heuristic, meta-heuristic or hybrid scheme)\cite{kumar_comprehensive_2019}\cite{singh_review_2017}\cite{houssein_task_2021}\cite{masdari_towards_2016}, by centralized or distributed methods\cite{avan_state---art_2023} or by application, technique, and metrics\cite{arunarani_task_2019}. However, these studies are limited to a specific scheduling environment. Although \cite{luo_resource_2021}~\cite{ghafari_task_2022}~\cite{alkhanak_cost_2016} summarize scheduling methods across two or more distributed environments (e.g., cloud and grid environment), none comprehensively covers research on all types of geo-distributed computing environments, especially scheduling in super computer (grid) environment. This paper aims to fill this gap by summarizing the diverse geo-distributed scheduling strategies across four specific computing environments: geo-distributed cloud, cloud-edge, edge, and geo-distributed supercomputer computing. We include HPC environment as it focuses on extreme performance optimization and large-scale resource utilization, fundamentally differing from other geo-distributed computing paradigms.

The main contribution of this paper is twofold. First, we investigate the latest research advancements in geo-distributed task scheduling, classifying relevant works according to their scheduling environments. Second, we dive into each environment and classify the works based on three goals: performance, fault-tolerance, and fairness. Performance ensures efficient resource utilization and minimizes costs, fault-tolerance guarantees system reliability in failure-prone distributed systems, and fairness focuses on equitable resource allocation in multi-tenant settings. Within performance, we further explore methods targeting computing resource utilization, such as heuristic, AI-based, mathematical, and hybrid approaches. Hybrid methods, such as AI combined with heuristic techniques, leverage the strengths of multiple paradigms. While computing resource utilization emphasizes efficient use of hardware, data transfer efficiency aims at storage and network optimizations to minimize latency and enhance I/O performance.

Fig. \ref{fig:Overview_of_geo-distributed_task_scheduling} illustrates the organization of the remainder of this survey. Section 2 discusses task scheduling techniques in the geo-distributed cloud computing environment. Section 3 covers scheduling techniques in the edge environment. Section 4 introduces task scheduling techniques in the cloud-edge environment. Section 5 examines task scheduling strategies in the geo-distributed supercomputer computing environment, with a uniform classification across each environment by scheduling goals: fairness and fault-tolerance. Section 6 discusses the opportunities and challenges of task scheduling in geo-distributed computing. Finally, we conclude this survey in Section 7.

\section{Geo-Distributed Cloud Computing}

Geo-distributed cloud computing (GDCC) operates across data centers (DCs) situated in diverse geo-locations, characterized by preemptible resources in a multi-tenant environment, infrastructure heterogeneity across regions and elasticity in resource scaling. This architecture presents multiple challenges, including inter-DC network latency, bandwidth constraints, and regional regulatory compliance requirements. This often involves scenarios with multiple cloud service providers, adding complexity due to the diversity of cloud environments. Scheduling systems navigate varying pricing models, resource allocation policies, and different regulations for each provider, while addressing provider-specific API limitations and cross-provider communication requirements. Task scheduling in GDCC addresses these challenges while optimizing resource utilization, minimizing latency, and reducing operational costs across geo-distributed DCs. (e.g., employing data locality for enhanced I/O performance). Fig. ~\ref{fig:Studies_on_computation_optimizations_for_distributed_LLM_training.} summarizes scheduling strategies addressing these geo-specific challenges in GDCC systems.


\tikzstyle{boxStyle}=[
    rectangle,
    draw=black,
    rounded corners,
    text opacity=1,
    minimum height=2em,
    minimum width=5em,
    inner sep=2pt,
    align=center,
    fill opacity=.5,
    line width=0.5pt,
]
\tikzstyle{leaf}=[boxStyle, minimum height=1.5em, minimum width=28em,
    fill=gray!15, text=black, align=left,font=\normalsize,
    inner xsep=2pt,
    inner ysep=4pt,
    line width=0.5pt,
]

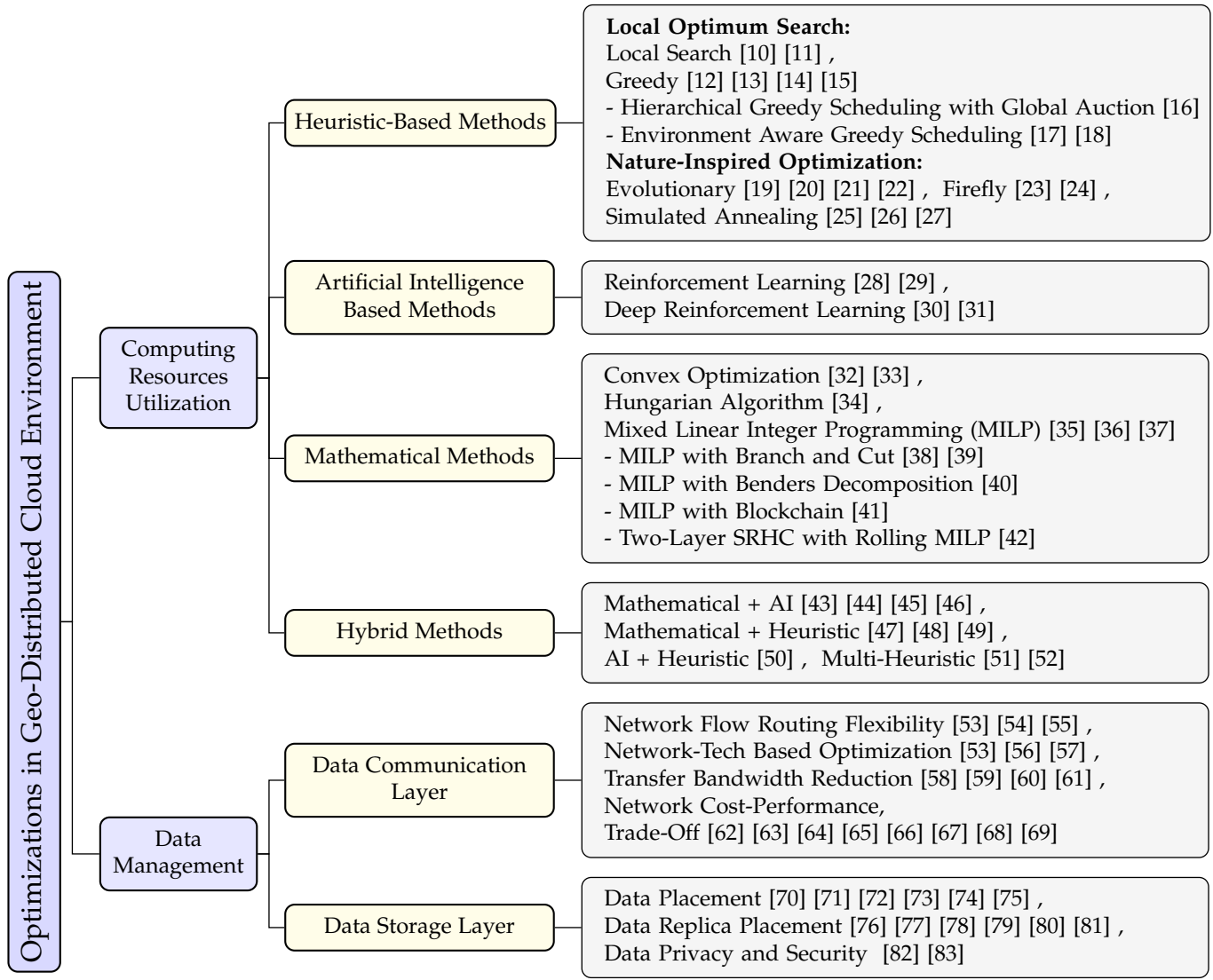
\begin{figure*}[htpb]
    \centering
    \resizebox{\textwidth}{!}{
        \begin{forest}
        forked edges,
        for tree={
            grow=east,
            reversed=true,
            anchor=base west,
            parent anchor=east,
            child anchor=west,
            base=center,
            font=\large,
            rectangle,
            draw=black,
            rounded corners,
            align=left,
            text centered,
            minimum width=4em,
            edge+={black, line width=0.5pt},
            s sep=8pt,   
            inner xsep=4pt,
            inner ysep=4pt,
            line width=0.8pt,
            sub_nodes/.style={rotate = 90, child anchor=north, parent anchor=south, anchor=center},
        },
        where level=1{minimum width=7em,font=\normalsize,}{},
        where level=2{minimum width=12em,font=\normalsize,}{},
        [
        {Optimizations in Geo-Distributed Cloud Environment}, sub_nodes, fill=blue!15
                [
                {Computing\\ Resources\\ Utilization}, fill=blue!10, align=center
                    [
                    {Heuristic-Based Methods}, fill=yellow!10
                        [
                        \textbf{Local Optimum Search:} \\
                        Local Search~\cite{li_energy-aware_2022}~\cite{hussain_deadline-constrained_2022} {, } \\Greedy~\cite{wang_energy_2022}~\cite{li_efficient_2022}~\cite{li_cluster_2021}~\cite{chen_big_2020} \\
                        - Hierarchical Greedy Scheduling with Global Auction~\cite{choudhury_mast_nodate}  \\
                        - Environment Aware Greedy Scheduling~\cite{padhi_uncertainty_2024}~\cite{ali_spatial_2024} \\
                        \textbf{Nature-Inspired Optimization:} \\ 
                        Evolutionary~\cite{wu_multi-objective_2024}~\cite{ebadifard_federated_2021}~\cite{khalid_dual_2023}~\cite{yuan_revenue_2021} {, }
                        Firefly~\cite{nithyanantham_resource_2021}~\cite{ammari_firefly_2022} {, } \\
                        Simulated Annealing~\cite{yuan_energy_2022}~\cite{yuan_biobjective_2021}~\cite{yuan_energy-efficient_2022}
                        ,
                        leaf, 
                        text width= 26em
                        ]
                    ]
                    [
                    {Artificial Intelligence\\ Based Methods}, fill=yellow!10, align=center
                        [
                        Reinforcement Learning~\cite{zhou_adaptive_2022}~\cite{zhang_sustainable_2023} {, } \\
                        Deep Reinforcement Learning~\cite{bi_cost-optimized_2022}~\cite{zhao_deep_2021}
                        ,
                        leaf, 
                        text width= 26em
                        ]
                    ]
                    [
                    {Mathematical Methods}, fill=yellow!10
                        [
                        Convex Optimization~\cite{yuan_geography-aware_2022}~\cite{kiani_profit_2018} {, } \\
                        Hungarian Algorithm~\cite{li_mapreduce_2022} {, } \\
                        Mixed Linear Integer Programming (MILP)~\cite{hao_joint_2024}~\cite{wang_stochastic_2020}~\cite{wang_optimal_2018}  \\
                        - MILP with Branch and Cut~\cite{souza_casper_2024}~\cite{zhao_workload_2023} \\
                        - MILP with Benders Decomposition~\cite{yang_carbon_2023} \\
                        - MILP with Blockchain~\cite{sajid_blockchain-based_2021} \\
                        - Two-Layer SRHC with Rolling MILP~\cite{cao_managing_2024} 
                        ,
                        leaf, 
                        text width= 26em
                        ]
                    ]
                    [
                    {Hybrid Methods}, fill=yellow!10
                        [
                        Mathematical + AI~\cite{qin_joint_2021}~\cite{wang_turbo_2020}~\cite{hogade_energy_2022}~\cite{hogade_game-theoretic_2024} {, } \\
                        Mathematical + Heuristic~\cite{hosseinalipour_power-aware_2020}~\cite{li_effective_2020}~\cite{yuan_spatiotemporal_2019} {, } \\
                        AI + Heuristic~\cite{niu_gmta_2020} {, } 
                        Multi-Heuristic~\cite{yuan_profit-sensitive_2020}~\cite{yuan_fine-grained_2020} \\
                        ,
                        leaf, 
                        text width= 26em
                        ]
                    ]
                ]
                [
                {Data\\ Management}, fill=blue!10, align=center
                    [
                    {Data Communication\\Layer}, fill=yellow!10, align=center
                        [
                        Network Flow Routing Flexibility~\cite{chen_network_optimizing_2022}~\cite{li_endpoint-flexible_2020}~\cite{zhao_optimizing_2020} {, } \\
                        Network-Tech Based Optimization~\cite{chen_network_optimizing_2022}~\cite{mostafaei_sdn-enabled_2023}~\cite{gu_service_2020} {, } \\
                        Transfer Bandwidth Reduction~\cite{MaxCompute_Website}~\cite{huang_yugong_2019}~\cite{zhou_cost-aware_2020}~\cite{liu_job_2020} {, } \\
                        Network Cost-Performance{, } \\Trade-Off~\cite{xu_trading_2022}~\cite{oh_network_2022}~\cite{pradhan_optimal_2024}~\cite{fan_online_2024}~\cite{yang_less_2022}~\cite{tao_congestion-aware_2022}~\cite{song_hcec_2024}~\cite{marzuni_cross-mapreduce_2021}
                        ,
                        leaf, 
                        text width= 26em
                        ]
                    ]
                    [
                    {Data Storage Layer}, fill=yellow!10
                        [
                        Data Placement~\cite{li_optimal_2022}~\cite{xie_multi-objective_2022}~\cite{atrey_spech_2019}~\cite{wang_geocol_2021}~\cite{li_placement_2023}~\cite{convolbo_geodis_2018} {, } \\
                        Data Replica Placement~\cite{li_data_2019}~\cite{liu_scalable_2020}~\cite{yu_framework_2020}~\cite{emara_distributed_2020}~\cite{chen_qos-aware_2021}~\cite{liu_efficient_2019} {, } \\
                        Data Privacy and Security ~\cite{nithyanantham_hybrid_2022}~\cite{zhou_privacy_2019}
                        ,
                        leaf, 
                        text width= 26em
                        ]
                    ]
                ]
        ]
        \end{forest}
    }
    
    \caption{Taxonomy of studies on optimizations under geo-distributed cloud computing infrastructure.}
    \label{fig:Studies_on_computation_optimizations_for_distributed_LLM_training.}
\end{figure*}

\subsection{Performance}

\subsubsection{Computing Resources Utilization}

\textit{I. Heuristic-based Methods}

\textit{1) Local Optimum: }
Local optimum refers to a solution to an optimization problem where, within a neighboring set of candidate solutions, no better solution exists. Unlike a global optimum, a local optimum may not be the best possible solution overall, but it is the best in its immediate vicinity. 

\begin{figure}[t]
    \includegraphics[width=0.49\textwidth, page=1]{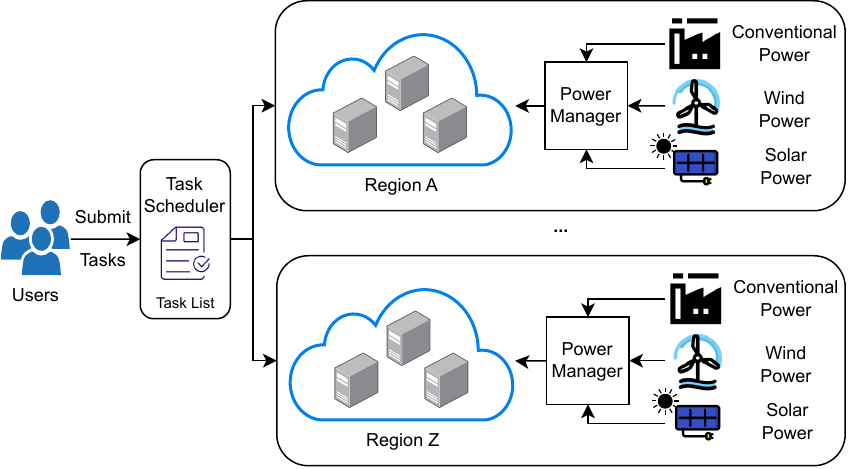}
    \caption{An example of a geo-distributed computing architecture exploiting spatial-temporal diversity. Every geo-distributed region has its own power supply and power price. After tasks are submitted, the scheduler will assign tasks to or offload tasks from certain computing regions considering each region's power supply diversity.}
    \label{fig:An_example_of_geo-distributed_computing_architecture_exploiting_spatial-temporal_diversity.}
\end{figure}

\textbf{Local Search Algorithms.}  
Fig.~\ref{fig:An_example_of_geo-distributed_computing_architecture_exploiting_spatial-temporal_diversity.} demonstrates how electricity prices vary across different time periods and geographical locations, enabling task schedulers to allocate computing tasks appropriately to reduce electricity costs.
Li et al.\cite{li_energy-aware_2022} propose an energy-aware workflow scheduling method that considers inter-DC data transmission costs and regional electricity price dynamics. It reverses traditional Adaptive Local Search (ALS) by dynamically decreasing the number of swapped immediate successor task pairs in each neighborhood iteration.
In contrast, DEWS (Deadline-constrained Energy-aware Workflow Scheduling) algorithm \cite{hussain_deadline-constrained_2022} employs Variable Neighborhood Descent (VND) to swap in-layer tasks and select geo-distributed DCs through three neighborhood structures. It then integrates Dynamic Voltage and Frequency Scaling (DVFS)-based energy optimization to adjust VM frequency and fully utilize task slack time.

\textbf{Greedy Algorithms. }For heterogeneous geo-distributed MapReduce clusters, Wang et al.\cite{wang_energy_2022} propose a three-phase dynamic scheduling framework that prioritizes data locality by scheduling tasks to the nearest available servers (rack-local, cluster-local, or remote). 
In geo-distributed cloud DCs, job execution can be hindered by stragglers, including both tasks and nodes.
To deal with straggling nodes, Li et al.\cite{{li_efficient_2022}} first detect them through statistical analysis of historical performance metrics, including authority category, urgency and length. Then it maps tasks to resources through a priority-based, time-cost trade-off calculation for optimal resource utilization. This explicitly prevents task assignment to these identified straggling nodes while redistributing their existing tasks to normal nodes with available capacity.
To deal with straggling tasks, Li et al.\cite{li_cluster_2021} propose a two-phase speculative execution strategy that selects nodes with the strongest processing capability to create task replicas. First, it evaluates cluster load to identify straggler-affected jobs; then, it greedily chooses nodes with the highest computing power, storage capacity and memory resources to execute task replicas. This greedy node selection ensures the replicas can be processed as quickly as possible to mitigate the impact of stragglers.
Similarly, Real-Time Scheduling Algorithm using Task Duplication (RTSATD)\cite{chen_big_2020} focuses on big data processing workflows by selecting tasks with minimum earliest start time (MESTF) while duplicating precursor tasks to the same instance in geo-distributed clouds.

\textbf{Hierarchical Greedy Scheduling with Global Auction.} MAST (ML Application Scheduler on Twine) \cite{choudhury_mast_nodate} introduces a three-level hierarchical scheduler that decouples traditional monolithic cluster scheduling into three hierarchical scopes, including global queue management, regional resource allocation, and cluster-level orchestration, where jobs are scheduled through a distributed auction mechanism. Regional ML Schedulers compete to host workloads by calculating placement quality scores based on resource availability and preemption cost, enabling exhaustive evaluation across regions before making final placement decisions.

\textbf{Environment Aware Greedy Scheduling.} Using renewable energy not only reduces energy costs, but also is more environmentally friendly. But renewable energy supply is often unstable and varying constantly.
Based on uncertainty level (UNL) of renewable energy, Padhi et al.\cite{padhi_uncertainty_2024} develop four scheduling algorithms based on UNL to optimize energy allocation using variable renewable and non-renewable energy sources. UNL categorizes uncertainty from low to high for users and from 1\% to 100\% for cloud providers, forming the basis for the following algorithms: \textit{UNL-FABEF} reduces operational costs by optimizing energy usage predictions; \textit{UNL-HAREF} maximizes renewable energy utilization and minimizes carbon emissions; \textit{UNL-RR} evenly distributes tasks among DCs in a cyclical manner; and \textit{UNL-MOSA} is a hybrid approach that dynamically adapts to changes in energy availability for efficient resource utilization and cost-effectiveness.
By considering computer room air condition (CRAC) operations with workload scheduling, Ali et al.\cite{ali_spatial_2024} propose spatio-thermal-aware workload management algorithms that always select the lowest-cost DC from a sorted list based on cooling efficiency and electricity prices, while considering temperature variations (inside/outside DCs). These approaches use a zone-based (cool, warm, hot) allocation scheme to greedily select servers with minimum cooling requirements, reducing both cooling costs and service level agreement (SLA) violations in geo-distributed environment.

\textit{2) Nature-Inspired Algorithms: } This type of algorithm mimics processes and behaviors observed in nature, which are commonly used to solve complex optimization problems by exploring large search spaces and avoiding local optima through mechanisms.

\textbf{Evolutionary Algorithms. } 
For geo-distributed cloud computing environment, researchers have proposed different evolutionary approaches to handle the complexity of task scheduling with multiple objectives and data locality constraints. Wu et al.\cite{wu_multi-objective_2024} propose a data locality-aware multi-workflow scheduling mechanism for federated clouds that first pre-processes tasks sharing same datasets to reduce data transfer volume, then uses evolutionary multi-objective optimization with intensification strategy to minimize both makespan and rental costs while meeting deadline constraints.
More recently, Ebadifard et al.\cite{ebadifard_federated_2021} address seven conflicting objectives scheduling through an enhanced Grid-based Evolutionary Algorithm (GrEA). GrEA first partitions computation/data-intensive tasks using hierarchical clustering for data locality, then applies a $\theta$-dominance relation that reduces hyper-volume computation from exponential to $O(n^2)$ complexity while maintaining solution diversity through grid dominance, data normalization and reference point optimization.
Focusing on energy costs in geo-distributed clouds, Khalid et al.\cite{khalid_dual_2023} formulate this as a constrained bi-objective optimization problem and leverage the Strength Pareto Evolutionary Algorithm (SPEA-II) to iteratively determine Pareto-optimal solutions for request dispatch and resource allocation, considering both computing and cooling costs under smart grid dynamics.
Taking the advantage of spatial variations, Yuan et al.~\cite{yuan_revenue_2021} propose an improved multi-objective evolutionary algorithm based on decomposition (IMEAD) decomposing the revenue-energy cost optimization problem into multiple sub-problems. Then it evolves solutions through genetic operators to determine optimal task splitting ratios and service rates under renewable energy constraints.

\textbf{Firefly Algorithms.} Firefly algorithms are an optimization technique where solutions ``attract'' better ones, mimicking the behavior of fireflies.
Handling geographically distributed large data with resource and cost optimization is a key challenge.
Nithyanantham et al.\cite{nithyanantham_resource_2021} introduce a Multivariate Metaphor based Metaheuristic Glowworm Swarm Map-Reduce Optimization (MM-MGSMO) technique which uses virtual machines (glowworms) and update their positions based on multiple objective functions including bandwidth, storage, energy and computation costs, followed by MapReduce-based allocation to optimize resource utilization and workload distribution.
In contrast, focusing on delay constraints and renewable energy utilization, Ammari et al.\cite{ammari_firefly_2022} address application scheduling in distributed Green DCs through a modified Firefly Algorithm (mFA) that dynamically adjusts attractiveness and introduces adaptive randomization parameter with damping to maximize renewable energy usage across geographical locations.

\textbf{Simulated Annealing (SA) Algorithms. } SA algorithms are optimization methods that mimic the metal cooling process, gradually refining solutions to reach the global optimum. 
Yuan et al.\cite{yuan_energy_2022} propose SA-based adaptive differential evolution (SADE) to balance task response time and energy cost in distributed DCs, which integrates Metropolis criterion and adaptive mutation with entropy-based crowding distance for better convergence. For green cloud DCs, Yuan et al.\cite{yuan_biobjective_2021} develop Simulated-annealing-based biobjective differential evolution (SBDE) that uniquely optimizes both revenue and energy consumption by considering spatial variations in renewable power generation and electricity pricing. Targeting QoS in cloud environments, Yuan et al.\cite{yuan_energy-efficient_2022} present an adaptive bi-objective differential evolution (ASBD) that minimizes both energy cost and task loss probability through genetic operations and adaptive elite archive updates.
While sharing simulated annealing as their core optimization strategy, these methods differ in how they integrate SA with other techniques: SADE combines SA with differential evolution, SBDE incorporates SA into biobjective optimization, and ASBD adapts SA for elite archive-based evolution.

\vspace{\baselineskip}
\noindent \textit{II. AI-based Methods}

\textbf{Reinforcement Learning (RL).}
Graph partitioning is an important problem of graph analytics, involves analyzing large datasets spread in geo-distributed DCs.
RLCut\cite{zhou_adaptive_2022} is an adaptive graph partitioning method leveraging RL to obtain better performance and cost efficiency. It employs multi-agent learning to optimize hybrid-cut model decisions, considering both network bandwidth heterogeneity among distributed DCs and graph dynamicity to adaptively balance partitioning effectiveness and overhead.

Scheduling AI-Generated Content (AIGC) workloads in the global cloud system needs to consider special characteristics of ML training, such as gang scheduling, locality of GPUs, intensive and exclusive GPU usage.
Zhang et al.'s\cite{zhang_sustainable_2023} algorithm leverages the advantages of multi-agent reinforcement learning (MARL) and Soft Actor Critic (SAC) algorithms to optimize GPU utilization while minimizing operational costs and carbon emissions. MARL eliminates the single point of failure in the central scheduling system and is scalable when the network grows, while SAC balances policy exploitation with action exploration optimally and has the advantage of addressing complex reward structures such as delayed rewards. 

\textbf{Deep Reinforcement Learning (DRL). } Due to the uncertainty and complexity of energy availability and task arrival in green DCs, traditional heuristic algorithms encounter difficulties in geo-distributed task scheduling and resource allocation. Bi et al.\cite{bi_cost-optimized_2022} introduce an Improved Deep Q-learning Network (IDQN) that enables an agent to learn from a reward function and continuously select optimal green DCs and servers to maximize the reward, resulting in lower task rejection rates and energy costs.
Facing the same problem, Zhao et al.\cite{zhao_deep_2021} propose a Proximal Policy Optimization based DRL approach, which automatically applies workload shifting and cloud-bursting in a hybrid multi-cloud environment consists of multiple private and public clouds to maximize renewable energy utilization and avoid deadline constraint violations.

\vspace{\baselineskip}
\noindent \textit{III. Mathematical Methods}

\textbf{Convex Optimization.} This is one of the mathematical approaches where the objective function is convex, meaning any local minimum is also a global minimum, ensuring efficient problem-solving.
Considering spatial cost and revenue variations of distributed green DCs, Yuan et al.\cite{yuan_geography-aware_2022} formulate a profit maximization problem as a convex optimization and address with their Geography-Aware Task Scheduling (GATS) approach using the Interior Point Method.
Kiani and Ansari\cite{kiani_profit_2018} propose a profit-maximizing workload distribution strategy for workload distribution across geo-dispersed green DCs. It decomposes workloads into green and brown components served by renewable and traditional energy sources respectively, optimizing both workload allocation and service rates while accounting for SLAs and electricity price diversity across regions. The strategy leverages a G/D/1 queuing model to capture workload distribution and proves the convexity of the optimization problem.

\textbf{Hungarian Algorithm.} Li et al.\cite{li_mapreduce_2022} propose a MapReduce scheduling framework optimizing both map and reduce phases: first matching map tasks to containers by considering both inter/intra-DC data locality costs, then assigning reduce tasks to geo-distributed nodes by optimizing cross-DC data transmission times and heterogeneous processing capabilities, while using heartbeat detection to maintain balanced resource utilization across the distributed infrastructure.

\textbf{Mixed Linear Integer Programming (MILP). } MILP models problems using linear equations while allowing discrete decision variables, enabling it to handle combinatorial complexity and ensure feasible solutions in scheduling tasks.

Hao et al.\cite{hao_joint_2024} propose a hybrid operation optimization to reduce both the electricity cost and carbon emission in geo-distributed DCs by jointly considering computational workload scheduling, carbon emission, micro grid operation and characteristics of Uninterruptible Power Supply (UPS). It utilizes the degree of freedom in computational workload scheduling to limit the nonlinear growth of UPS power losses and introduces carbon tax as a parameter in the optimization object.
Wang et al.\cite{wang_stochastic_2020} combine electrical and thermal system optimization in DC microgrids, which integrates scheduling with waste heat recovery, repurposing it for residential heating demands. By addressing the stochastic nature of renewable energy supply, delay-tolerant workloads, and thermal demand, their formulation minimizes total costs while ensuring system security, service quality and energy efficiency.
Wang et al.\cite{wang_optimal_2018} also formulate this as a MILP, incorporating QoS constraints modeled through an M/G/1 queuing network. But they transform it into a tractable form and propose a strategy powered by both renewable and conventional energy, incorporating dynamic voltage and frequency scaling.

\textbf{MILP with Branch and Cut. }
CASPER\cite{souza_casper_2024} is a carbon-aware scheduling and provisioning system for distributed web services. It formulates a multi-objective optimization problem utilizing spatial-temporal variability in energy sources and solves it using PuLP library, an interface to the Coin-or branch and cut (CBC) solver, to align computational workloads with available green energy across different regions.

\textbf{MILP with Branch and Bound. }
OPRS (Optimal Power Regulation Scheduling)\cite{zhao_workload_2023} optimizes power consumption by intelligently redistributing tasks based on demand response signals. It combines three power regulation methods, including task delay scheduling, hybrid cooling systems, and UPS utilization, to minimize total operating costs. Employing a branch-and-bound algorithm, OPRS addresses this issue as a MILP problem to achieve the balance between power reduction and performance.

\textbf{MILP with Benders Decomposition. }
To trade off between emission cutting effects from scheduling and carbon costs of workload migration, Yang et al.\cite{yang_carbon_2023} propose a large-scale MILP problem based on a spatio-temporal task migration mechanism and solve it using Benders decomposition algorithm which decouples task migration decisions and optical routing schemes across distributed DCs for carbon emission optimization.

\textbf{Blockchain-Enabled Distributed MILP. }
Sajid et al.\cite{sajid_blockchain-based_2021} design a decentralized energy-optimization system where DCs coordinate through a custom blockchain structure that enables direct workload migration based on real-time energy costs. Each DC employs MILP with conditional constraints to optimize across multiple energy sources (renewable/grid/battery/diesel) while using proof-of-work consensus to validate cost-based scheduling decisions. This framework replaces traditional front-end schedulers by enabling DCs to autonomously migrate workloads through blockchain-verified transactions when local energy costs exceed neighboring centers.

\textbf{Two-Layer SRHC with Rolling MILP.}
DCs often consume lots of electricity and thus can be used to balance the power market. Cao et al.\cite{cao_managing_2024} develop a two-layer Stochastic Receding Horizon Control (SRHC) optimization framework for managing DC clusters as non-wire alternatives: the upper layer optimizes market bidding through stochastic programming while the lower layer executes spatial-temporal workload scheduling through MILP. This framework recursively solves finite-horizon optimization problems to handle uncertainties in regulating prices and workload delays, enabling DCs to participate in power market balancing.

\vspace{\baselineskip}
\noindent \textit{IV. Hybrid Methods}

\textbf{Mathematical + AI. }
Qin et al.\cite{qin_joint_2021} leverage Lyapunov optimization to transform time-coupled carbon emission constraints into a queue stability problem for geographical load balancing, and then employs both Generalized Benders Decomposition (GBD) and Deep Q-Network (DQN) to optimize joint energy consumption across servers and network traffic in geo-distributed DCs.
Turbo\cite{wang_turbo_2020} is a geo-distributed analytics system that leverages LASSO and GBRT to predict query execution time and intermediate output sizes in real-time. It dynamically adjusts query plans based on resource fluctuations, seamlessly integrating with existing frameworks to enhance efficiency by reordering joins during execution.

Nash equilibrium-based Intelligent Load Distribution (NILD)\cite{hogade_energy_2022} combines game theory with Reinforcement Learning for workload management. This non-cooperative game-theoretic approach achieves optimal load balancing by simultaneously minimizing DC operational costs and response latency across geographical locations.
However, NILD does not consistently achieve global optima solutions.
Game-Theoretic Deep Reinforcement Learning (GT-DRL) \cite{hogade_game-theoretic_2024} advances carbon-aware scheduling by integrating location-specific renewable energy patterns into workload distribution across geo-distributed DCs. By synthesizing non-cooperative game theory with DRL, GT-DRL dynamically optimizes both carbon emissions and operational costs for AI inference workloads, adapting to real-time variations in electricity pricing and data transfer costs across geographical locations.

\textbf{Mathematical + Heuristic. }
Hosseinalipour et al.\cite{hosseinalipour_power-aware_2020} tackle energy optimization in geo-distributed DCs through a scale-adaptive framework for graph-structured tasks. Their approach combines convex programming for small-scale networks with cloud crawler-based sub-graph extraction for large-scale geo-distributed environments, while employing online learning mechanisms to adapt to dynamic pricing scenarios.
For distributed workflow scheduling, Li et al.\cite{li_effective_2020} advance the efficiency of cloud workflows with a hypergraph partitioning based scheduling strategy in geo-distributed DCs, which incorporates the cloud's state and utilizes the Dijkstra algorithm with a Fibonacci heap. The result is a significant reduction in both average task execution time and overall energy consumption, contributing to more balanced and sustainable cloud operations.
While the above work focuses on structural optimization, Yuan et al.\cite{yuan_spatiotemporal_2019} leverage spatial-temporal diversity in geo-distributed DCs and propose a spatial-temporal task scheduling (STTS) leveraging spatial-temporal diversity in geo-distributed DCs. By formulating energy cost minimization as a nonlinear constrained optimization problem, STTS combines genetic algorithms with simulated annealing and particle swarm optimization to achieve optimal task scheduling while considering geographical variations in both grid and renewable energy pricing.

\textbf{AI + Heuristic. }
The Geo-aware Multi-Agent Task Allocation (GMTA) \cite{niu_gmta_2020} framework leverages multi-agent auction mechanisms to optimize scientific workflow execution across geo-distributed container-based clouds. GMTA enhances parallel execution while simplifying dependencies by intelligent workflow partitioning and agent-based negotiation.

\textbf{Multi-Heuristic. }
Profit-sensitive spatial scheduling (PS3)\cite{yuan_profit-sensitive_2020} uses a genetic-simulated-annealing-based particle swarm optimization, which leverages spatial factors such as revenue, power grid price, solar radiation, wind speed, energy capacity, and air density to maximize the total profit of a geo-distributed green DC (GDGDC) provider while meeting task response time constraints.
Under the same environment, the Simulated-annealing-based Bees algorithm (SBA) \cite{yuan_fine-grained_2020} tackles fine-grained scheduling challenges through an queuing-theoretic approach. By leveraging a G/G/1 queuing model, SBA addresses the geographical variations in power pricing and green energy availability across different DC locations. It simultaneously optimizes three key aspects: workload distribution patterns, server operating speeds, and the number of active servers at each geographical location, while maintaining strict response time requirements.

\subsubsection{Data Management}

\noindent \textit{I. Data Communication Layer}

Efficiently and cost-effectively accessing the required data with low latency for geographically distributed computing tasks is essential, especially when the required data is distributed across various locations with limited cross-domain transfer bandwidth.

\begin{figure}
    \includegraphics[width=0.49\textwidth, page=1]{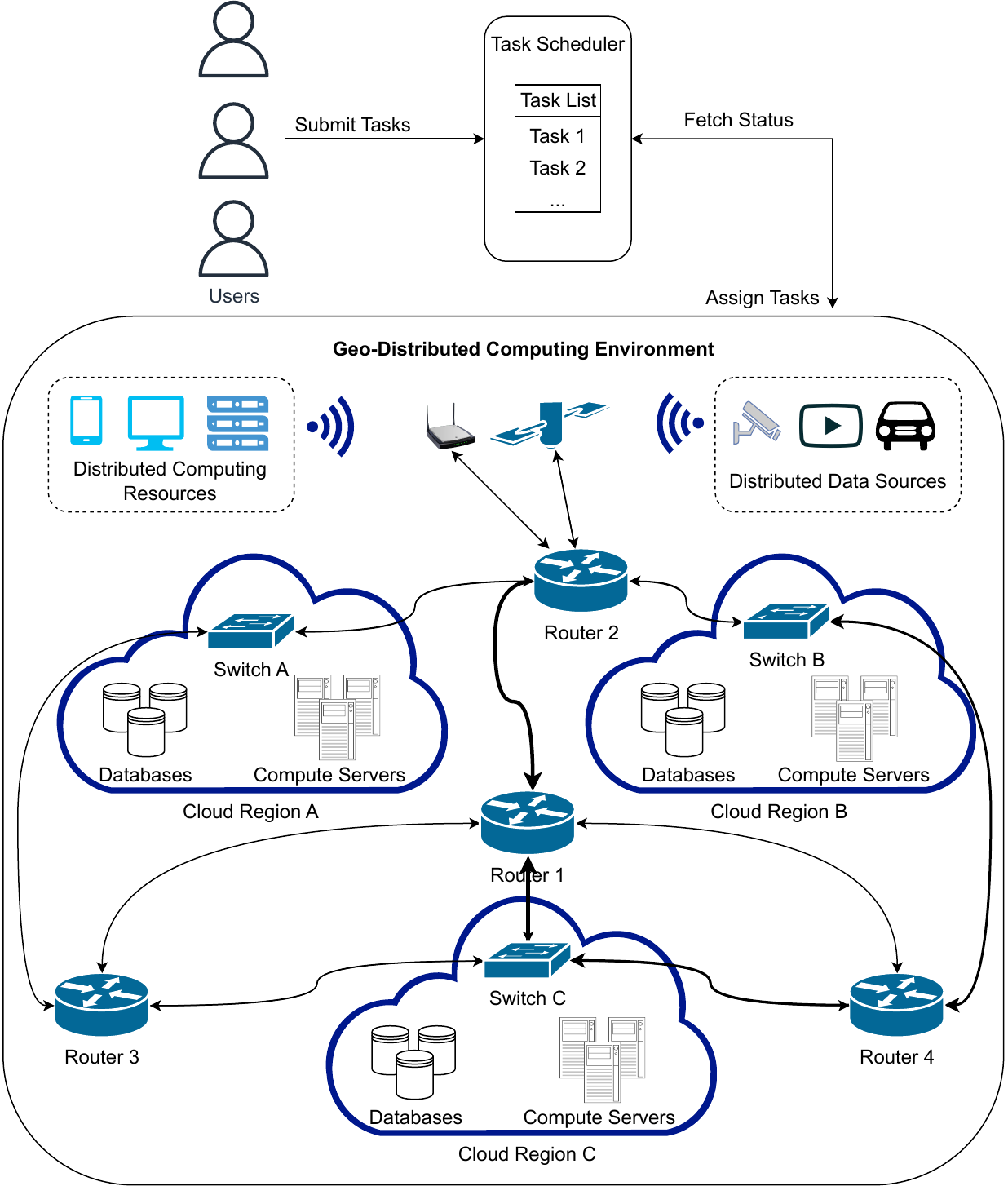}
    \caption{An example of a geo-distributed computing environment focusing on distributed network architecture. After users submit tasks, the scheduler will assign tasks to different computing nodes according to computing and networking status. The thickness of the lines between routers and switches represents the relative size of the bandwidth. The length of the lines indicates the relative transmission distances.}
    \label{fig:An_example_of_geo-distributed_computing_network_architecture.}
\end{figure}

\textbf{Network Flow Routing Flexibility.} Due to the vast differences in network topology and bandwidth among DCs, a flexible routing approach becomes crucial in mitigating congestion and enhancing network utilization. Intelligent network routing strategies ensure balanced utilization of links between DCs, facilitating efficient and equitable distribution of data transfer loads, thus speeding up application execution speed (see Fig.~\ref{fig:An_example_of_geo-distributed_computing_network_architecture.}).

Network flows will be generated to transfer the intermediate data between consecutive stages for further processing. These flows are collectively defined as a \textit{coflow} of the data analytic job.
Li et al.\cite{chen_network_optimizing_2022} propose a linear programming method to split and route data flows to multiple network paths and dynamically adjust sending rates to optimize bandwidth utilization across DCs. They treat the group of flows in a coflow that have the same pair of source and destination DCs as the basic unit in their multi-path routing model. 
For Map-Reduce jobs, in the shuffle phase, the entire set of network flows generated from map tasks to reduce tasks is referred to as a \textit{coflow}. Li et al.\cite{li_endpoint-flexible_2020} introduce Smart Coflow, which integrates endpoint flexibility into coflow scheduling, allowing for dynamic adjustment of data flow destinations based on current network conditions and DC availability.
HPS+\cite{zhao_optimizing_2020} uses an augmented hyper-graph model to represent task-data and data-DC dependencies. HPS+ applies hyper-graph partitioning to minimize WAN data transfers. Additionally, HPS+ introduces a Routing and Bandwidth Allocation (RBA) algorithm to coordinate data transfers and computation, prioritizing tasks with longer computing stages to reduce transfer times.

\textbf{Network Tech-based Optimization.} Network tech-based approaches primarily leverage SDN (Software Defined Network)'s routing control and VNF (Virtual Network Function)'s service flexibility to optimize geo-distributed data transfers.

\textit{SDN} is an architecture that allows for centralized control and dynamic management of network resources. Li et al.\cite{chen_network_optimizing_2022} provide a transfer optimization service for Spark, following the principle of SDN at the application layer, to fully control the routing for inter-DC traffic. For geo-distributed stream data analytics, Mostafaei et al.\cite{mostafaei_sdn-enabled_2023} also introduce a SDN-based framework that enables a SDN controller to monitor WAN conditions and dynamically select worker nodes based on network topology and link parameters. It integrates P4-based data plane implementation with network-aware scheduling, allowing efficient task allocation without modifying the underlying stream processing systems.
\textit{VNF} involves deploying network services as software instances rather than physical devices, allowing flexible network management. Gu et al.\cite{gu_service_2020} address the deployment of VNFs and network flow scheduling in distributed DCs to minimize the total cost of big data processing while ensuring QoS.

\textbf{Network Bandwidth Optimization.} MaxCompute\cite{MaxCompute_Website} is a fast, fully managed, TB/PB level data warehouse solution by Alibaba. It provides users with a comprehensive data import solution and a variety of classic distributed computation models, which can solve the problem of users' massive data computation faster, effectively reduce the cost of the enterprise, and ensure data security. 
Based on it, Huang et al.\cite{huang_yugong_2019} propose \textit{Yugong}, which works seamlessly with MaxCompute in very large scale production environments. By project migration, table replication and job outsourcing, the cross-DC bandwidth usage reduces significantly. 

Multi-level heterogeneities in network bandwidth and communication prices in geo-distributed DCs raises challenges to existing graph partitioning methods. To address it, Geo-Cut \cite{zhou_cost-aware_2020} first uses a cost-aware streaming heuristic to minimize inter-DC communication during edge assignment, followed by partition refinement to alleviate bottlenecks and optimize data transfer within budget constraints.

Geo-distributed machine learning (Geo-DML) applications also face challenges with limited WAN bandwidth and data privacy laws, hindering efficient model training across dispersed DCs. RoWAN\cite{liu_job_2020} (Routing and rate allocation in optical WAN) dynamically adjusts the network topology and allocates resources for each data flow. Additionally, they employ delayed SWRT (delayed Shortest Weighted Remaining Time) to prioritize and schedule multiple ML jobs effectively.

\textbf{Network Transfer Cost and Performance Trade-Off.} A crucial challenge in geo-distributed analytics (GDA) is efficiently managing the trade-off between cost and system performance. Xu et al.\cite{xu_trading_2022} address this challenge through a two-time scale approach that combines data placement optimization with query request admission control. Their method leverages Lyapunov optimization for effective online decision-making, enhancing both economic and operational efficiency without requiring future traffic predictions.
\textit{Kimchi}\cite{oh_network_2022} tackles heterogeneous data transfer costs by intelligently scheduling tasks based on network transfer costs, bandwidth availability, and data locations. This comprehensive approach significantly reduces operational expenses while maintaining query performance. Taking optimization a step further, GDA-OPT\cite{pradhan_optimal_2024} uniquely combines join order and job location optimization through dynamic programming. Its sophisticated cost model accounts for WAN costs, DC locations, and heterogeneous capabilities, while employing search space pruning techniques for efficient large-scale GDA management.
Geo-Distributed ML (Geo-DML) also meets with this problem, Training Flow Adaptive Steering (TFAS)\cite{fan_online_2024} is an online training flow scheduling algorithm for Geo-DML jobs over dynamic and heterogeneous WANs. They utilize a primal-dual framework within a linear programming model to optimize the allocation of network resources, expedite training completions and maximize ISP revenue.

Fulfilling all users’ request sometimes leads to high expenditure. Cloud providers can selectively accept user requests instead of fulfilling all to maximize service profits. Yang et al.\cite{yang_less_2022} propose dual solutions: \textit{Metis}, an offline algorithm that alternately maximizes the service revenue under given bandwidth and minimize the bandwidth cost under given requests. Cloud providers could dynamically adjust the bandwidth to purchase and the requests to accept. \textit{OSA}, an online scheduling algorithm evaluates the impact of scheduling the requests and make decisions in real time.

Network congestion management is another dimension of cost-performance optimization. CONA (CONgestion-Aware) \cite{tao_congestion-aware_2022} employs matrix-based traffic allocation and link grading strategies to maximize profit in geo-distributed transfers. While CONA addresses general network congestion, modern distributed DL requires more sophisticated approaches. HCEC (High-Convergence and Efficient-Communication) \cite{song_hcec_2024} advances this field by implementing dynamic rate adaptation and Adaptive Layer-wise Communication, optimizing both model convergence and communication efficiency across geo-distributed DCs.

For specialized MapReduce applications, Cross-MapReduce\cite{marzuni_cross-mapreduce_2021} introduces Gshuffling to minimize inter-cluster data transfer. This approach distinguishes between intra- and inter-cluster traffic, employing local shuffling and strategic global reducer selection through a Global Reduction Graph, thereby achieving efficient load balancing and reduced data transfer overhead.

\vspace{\baselineskip}
\noindent \textit{II. Data Storage Layer}

Efficiently storing the data required for computation is crucial, as data is indispensable when performing geographically distributed computing tasks.

\textbf{Data Placement Optimization.} Data placement optimization problem can be approached through algorithmic methods such as graph-based optimization, heuristic techniques, and machine learning frameworks.

Considering capacity limitations and load balancing, Li et al.\cite{li_optimal_2022} utilize the Floyd algorithm to solve the minimum bandwidth cost problem, a multi-source shortest path problem with a weighted directed graph. Then they transform the objective function to a LP problem and employ the Lagrangian relaxation method to obtain a data placement scheme.
Xie et al.\cite{xie_multi-objective_2022} convert this into a multi-dimensional knapsack problem and also employ Lagrangian relaxation method to solve, but they use ant colony optimization to further optimize the solution.
SpeCH (Spectral Clustering on Hypergraphs)\cite{atrey_spech_2019} scales hypergraph partitioning using spectral clustering.
SpectralApprox improves efficiency with low-rank matrix approximations, while SpectralDist distributes computations across machines to handle large workloads.
Data placement problem ca also be solved using reinforcement learning (RL). Wang et al.\cite{wang_geocol_2021} propose \textit{GeoCol}, a geo-distributed cloud storage system with low cost and latency using RL. It dynamically splits data requests into sub-requests sent to different DCs, using Seasonal Auto-regressive Integrated Moving Average (SARIMA) to predict latency and RL to determine the number and destination of sub-requests.

Li et al.\cite{li_placement_2023} address the challenge of optimizing parameter server (PS) placement for Geo-DML. They focus on enhancing communication efficiency by selecting the most suitable DC to serve as the PS based on minimizing communication costs, by developing an approximation algorithm utilizing the randomized rounding method.

GeoDis\cite{convolbo_geodis_2018} optimizes data-intensive job scheduling by balancing data locality and inter-DC transfers. Firstly, tasks are ordered based on their data size and shortest job first policy. Then tasks are assigned on DCs with the least load while considering network bandwidth.

\textbf{Data Replica Placement Optimization.} Data replica placement maintains data copies across distributed DCs, simultaneously reducing access latency, balancing system load, and improving overall efficiency. The key principle is prioritizing data locality, where tasks are preferentially assigned to DCs that host the majority of their required data.

Li et al.'s\cite{li_data_2019} research propose two algorithms. To reduce execution delay of non-node-locality tasks, the DLO-migrate algorithm fetches input data in advance using idle network bandwidth. To short job completion time and avoid unnecessary data transformation, DLO-predict algorithm predicts hotspots to periodically transfer hot files to multiple DCs.
Liu et al.\cite{liu_scalable_2020} propose a scalable and adaptive method through offline community discovery and online community adjustment methods. The offline scheme determines the replica placement solution based on average read or write rates, offering scalability with linear computational complexity and distributed implementation. The online scheme adaptively handles bursty data requests without completely overriding the existing replica placement.
With joint considerations of the data-node relationships and the associations of data groups, Yu and Pan\cite{yu_framework_2020} propose a hypergraph-based data placement framework without a relaxation. By introducing the iterative process of routing and replica placement, their method can be applied under replica scenario.
Emara et al.\cite{emara_distributed_2020} propose two data distribution strategies for data analysis: one without replication and one with replication. These strategies leverage the random sample partition data model to convert big data into sets of data blocks and distribute data blocks across DCs. The experimental results show that the strategy \textit{without replication}, some data blocks are required to download from the remote DCs to a central DC for approximate analysis of the big data as a whole. The main advantage of this strategy is to separate the storage level from the analysis level. For the strategy \textit{with replication}, the data in each DC forms as a random sample of the whole distributed data, as a sample of the data on each DC is enough to be representative of the whole distributed data.
Chen et al.'s\cite{chen_qos-aware_2021} method utilizes a golden division approach for Zipf-like replica distribution. They transform the challenge into a block-dependence tree construction problem and simplify it into a graph partitioning problem. Their approach minimizes network traffic and ensures QoS for data blocks in MapReduce applications. 

\textit{Metadata} has a critical impact on the efficiency of scientific workflow scheduling as it provides a global view of data location and enables task tracking during execution. Liu et al.\cite{liu_efficient_2019} use relational DBMS to manage hot metadata. They combine the hot metadata management strategies with three scheduling algorithms, OLB (Opportunistic Load Balancing), MCT (Minimum Completion Time) and DIM (Data-Intensive Multi-site task scheduling) to provide hot metadata management for multi-site task scheduling.

\textbf{Data Security and Privacy.} Data transfer across geo-distributed DCs creates complex challenges for data transmission and storage across multiple jurisdictional boundaries, particularly in ensuring security and compliance with diverse international regulations such as the European Union's General Data Protection Regulation (GDPR). 

Nithyanantham et al.\cite{nithyanantham_hybrid_2022} introduce a hybrid DL framework which uses a DNN enhanced with Siamese training to safeguard against secondary data inference, effectively preserving user privacy during feature extraction and classification tasks. Additionally, the framework employs Glow-worm Swarm Optimization (GSO) to fine-tune the hyper-parameters of the DNN, ensuring optimal performance across distributed environments like Hadoop. 
Considering the challenge of multi-level data privacy constraints, Zhou et al.\cite{zhou_privacy_2019} introduce a process mapping algorithm that integrates the communication matrix for application processes with the varying network performance metrics of DCs, enabling optimized mapping of processes to nodes. This strategic alignment not only complies with stringent data privacy laws but also maximizes the efficiency of data transmission across regions.

\subsection{Fairness}

Fairness-driven scheduling approaches aim to equitably allocate resources across tasks, often using optimization techniques to achieve balanced performance among competing jobs.

Chen et al.\cite{chen_scheduling_2019} focus on achieving \textit{max-min fairness} among jobs. They formulate it as a lexicographical minimization problem and leverage the totally uni-modular property of linear constraints. This enables the transformation of the problem into an equivalent sub LP problem formulation, which is efficiently solvable to ensure fairness across competing jobs. The sub problems can be solved by any LP solver and are guaranteed to have the same solution to the original problem.

\subsection{Fault-Tolerance}

Fault-tolerant scheduling techniques integrate redundancy and predictive maintenance to maintain reliability and optimize resource use, even under dynamic and large-scale conditions.

Li et al.\cite{li_fault-tolerant_2022} propose a fault-tolerant scheduling strategy which takes the task cloning, anomaly detection, energy consumption, and the task deadline into account. A replica policy based on the speculative execution model guarantees the fault tolerance of the geo-distributed clouds and obtain high performance of Spark. Then a scheduling strategy for containerized Spark clusters under a heterogeneous environment is proposed. 
Similarly, the two-level Approximate Dynamic Programming (ADP)\cite{zhang_two-level_2019} algorithm uses a virtualized monitoring model to predict server health, minimizing fault tolerance costs by avoiding unhealthy servers, and integrates RL to address the complexity of large state and action spaces.

\section{Edge Computing}

\begin{figure}[h]
    \includegraphics[width=0.51\textwidth, page=1]{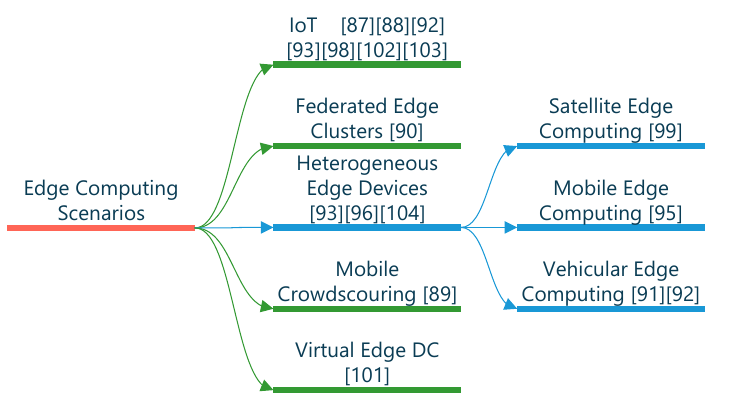}
    \caption{An overview of specific edge computing scenarios.}
    \label{fig:An_overview_of_computing_scenarios_in_edge_computing_environment}
\end{figure}

Edge computing is a geo-distributed computing paradigm that utilizes resources at the network edge to enable distributed computing near data sources (such as IoT devices and mobile devices). This distributed architecture effectively reduces communication latency, but it also introduces challenges: limited resources at edge nodes, unstable network connectivity, and high node heterogeneity. The scheduling process under edge environment typically involves monitoring available resources, assessing workload requirements, and making real-time decisions to allocate tasks to the most suitable edge nodes. It allocates workloads efficiently across edge nodes while considering resource limitations. It also ensures low latency by keeping tasks close to data sources. The following sections examine various scheduling approaches, each addressing specific edge computing challenges under specific edge computing scenarios (see Fig.~\ref{fig:An_overview_of_computing_scenarios_in_edge_computing_environment}).

\subsection{Performance}
\subsubsection{Computing Resources Utilization}

\textit{I. Heuristic-based Methods}

Many large-scale IoT applications need to analyze data distributed across multiple sites to obtain final results. The problem is how to efficiently execute tasks among edge nodes and devices, considering the heterogeneity of resource capacities and prices across multiple sites to ensure jobs finish before their deadlines.
 
\textbf{Gradient-based.} Chen et al.\cite{chen_geo-distributed_2022} characterize this as a deadline constrained quadratic programming problem and introduce a minimize the job completion cost before a given deadline (MCGL) method leveraging the negative correlation relationship between job completion time and job completion cost to solve.

\textbf{Distance-based.} Decomposable aggregation functions (DAFs) are distributed and parallelized across multiple compute nodes in stream processing engines to handle large IoT data. To efficiently deploy DAFs on resource-constrained distributed nodes, Chatziliadis et al.\cite{chatziliadis_efficient_2024} introduce NEMO, leveraging Euclidean embeddings of network topologies along with a set of heuristics to manage millions of nodes. It dynamically adjusts to topological changes through adaptive replacement and replication decisions.

\begin{figure}[h]
    \centering
    \includegraphics[width=0.35\textwidth, page=1]{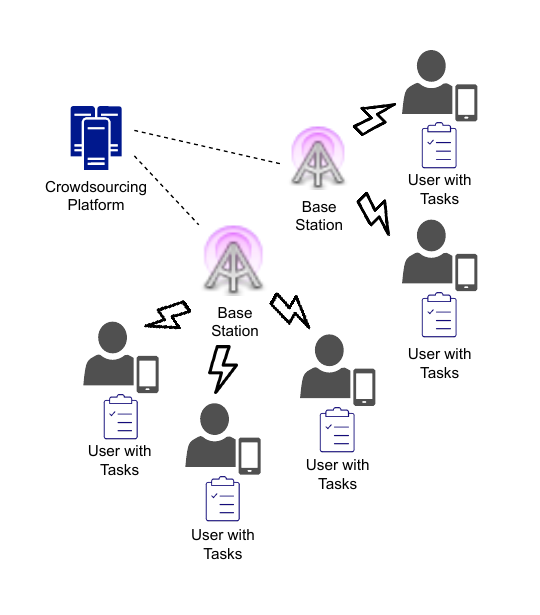}
    \caption{An overview of the Mobile Crowd-sourcing System (MCS).}
    \label{fig:An_overview_of_a_Mobile_Crowdsourcing_System (MCS).}
\end{figure}

\textbf{Divide-and-Conquer.} Mobile crowdsourcing leverages the collective efforts of individuals using mobile devices to gather data, complete tasks, and solve problems, often as part of IoT environments. An overview of such a system is shown in Fig.~\ref{fig:An_overview_of_a_Mobile_Crowdsourcing_System (MCS).}. Wang et al.\cite{wang_compact_2020} propose two approaches: a breath-first search-based dynamic priority algorithm for local optimization and an evolutionary multitasking algorithm for global optimization. The local optimization adopts a layered model and utilizes a divide-and-conquer technique to construct scheduling solution sequentially. The second tackles global optimization by solving multiple problems simultaneously, enhancing efficiency through knowledge transfer and collaborative information sharing between tasks.

\textbf{Affinity-based.} Microservice is a software architecture style where a complex application is broken down into small and independently deployable services, each focusing on a specific function and communicating over the network. Many large-scale application development patterns are moving towards agile microservice approach. Phare\cite{castellano_scheduling_2024}, based on affinity, prioritizes microservices with the more stringent requirements and places them on the most convenient computing facilities. 

\vspace{\baselineskip}
\noindent \textit{II. AI-based Methods}

\textbf{Deep Reinforcement Learning (DRL).} Liu et al.~\cite{liu_asynchronous_2023} propose a multi-resource orchestration framework in vehicular edge computing (VEC) that combines multi-hop Vehicle-to-Vehicle (V2V) offloading and service-migration-based Vehicle-to-Infrastructure (V2I) offloading. They employ an A3C algorithm where multiple worker agents learn optimal task scheduling policies through actor-critic networks. In heterogeneous IoT scenarios, Ren et al.'s~\cite{ren_collaborative_2024} propose a framework with macro (nBSCS) and micro (lBSCS) base station spaces. They deploy decentralized DRL agents at each base station to optimize offloading strategies based on available computing and caching resources.

\textbf{Multi Agent-based.} Tang et al.\cite{tang_distributed_2022} propose a distributed task scheduling framework for serverless edge computing in IoT. First the problem is formulated as a partially observable stochastic game, with each serverless edge node optimizing its own utility based on local observations. Then a dueling double deep recurrent Q-network (D3RQN) algorithm is applied, enabling each edge node to approximate optimal scheduling decisions without global information.

\vspace{\baselineskip}
\noindent \textit{III. Mathematical Methods}

\textbf{Mixed Integer Non-Linear Programming (MINLP).} Li et al.\cite{li_qos_2020} address task offloading in mobile edge computing with a focus on statistically guaranteed QoS to manage dynamic wireless conditions. The authors develop a statistical computation and transmission model as a MINLP with delay constraints and then leverage convex optimization and Gibbs sampling to balance task offloading and resource allocation.

\textbf{Quadratic and Dynamic Programming.} Michailidou et al.\cite{michailidou_optimizing_2024} propose a three-objective task allocation in multi-query edge analytics targeting latency, resource consumption, and Quality of Results (QoR). It combines quadratic and dynamic programming for task placement and data down-sampling, with adaptive techniques to revise allocations for new queries, optimizing resource usage.

\textbf{Modified Kuhn-Munkres.} Geo-distributed edges handle tasks offloaded from cloud DCs, but high energy costs burden service providers. Liao et al.\cite{liao_ev-assisted_2024} propose an electric vehicle (EV)-assisted edge computing architecture that leverages EVs' idle computing resources and stores energy charged during off-peak hours. Their solution incorporates a spatiotemporal workload offloading model that discretizes the optimization problem into smaller sub-problems in both time and space dimensions, and deploys a modified Kuhn-Munkres algorithm for dynamic matching between EVs and service requests based on energy costs and QoS constraints.

\vspace{\baselineskip}
\noindent \textit{IV. Hybrid Methods}

\textbf{Mathematical + Heuristic.} Rossi et al.\cite{rossi_elastic_2019} use an Integer Linear Programming formulation and a network-aware greedy heuristic for container-based application deployment. It selects the hosting VMs from a sorted list using a greedy approach. The list is sorted in ascending order, using the objective function as distance metric, the first VMs of the list minimizes the adaptation time.

\subsubsection{Data Management}

The network and storage layers' scheduling algorithms play essential roles in enhancing data transfer efficiency across distributed and heterogeneous edge systems. The network layer optimizes data flow scheduling for distributed and satellite edge computing, while the storage layer enhances data aggregation and manage metadata to ensure efficient, low-latency access in geo-distributed environments.

\vspace{\baselineskip}
\noindent \textit{I. Data Communication Layer}

\textbf{Online Algorithms.}
\textit{Okita} and \textit{Okita*} \cite{pang_online_2022} are two online scheduling algorithms.
\textit{Okita} determines both worker and parameter server placement across edge sites to minimize network bandwidth usage, while \textit{Okita*} employs a non-preemptive fashion and optimizes this further by using dynamic programming to divide training data into time slots, making scheduling decisions based on data locality and wireless resource constraints.

\textbf{Satellite Edge Computing (SEC) Network.} Satellites, equipped with computing resource, have been envisioned as a key enabling technology to timely analyze stream data of IoT applications in remote regions on the earth. Streaming analytics, leveraging SEC within terrestrial-satellite networks, enables timely processing of large IoT data streams in remote regions by using satellites equipped with computing resources.
Xu et al. \cite{xu_enabling_2024} address the flow time minimization problem in SEC for big data analytics by formulating it as an Integer Linear Programming (ILP) problem. They propose an offline approximation algorithm based on auxiliary graph construction and an online learning algorithm with bounded regret, leveraging Lipschitz bandit techniques to handle the dynamic movement of satellites and uncertain dataset volumes.

\textbf{Edge Compute First Networking (ECFN).} ECFN integrates edge computing with networks to enable efficient data processing. Liu et al.\cite{liu_multi-stage_2023} divide data processing into multiple parallel stages, where each stage optimizes cluster center selection and light-path provisioning to minimize job completion time. They further develop a routing and frequency slot reallocation scheme based on stage completion time to reduce bandwidth consumption during data transmission.

\vspace{\baselineskip}
\noindent \textit{II. Data Storage Layer}

\textbf{Metadata Management.} Metadata Management systematically organizes and maintains the descriptive information and attributes about the stored data, which is crucial for enabling efficient data access, retrieval, and management across the distributed storage infrastructure. Dou et al.\cite{dou_architecture_2022} introduce a virtual edge DC with intelligent metadata service, which dynamically aggregates idle storage capabilities and divides the file system directory tree to efficiently manage metadata in distributed file systems.

\subsection{Fairness}

Fairness-focused scheduling methods address equitable resource distribution, ensuring that all users or tasks receive proportionate access to edge computing resources.

\textbf{Dynamic Nash Bargaining Game.} FairHealth\cite{lin_fairhealth_2022}, a 5G edge healthcare scheme that ensures long-term proportional fairness in the Internet of Medical Things by addressing priority-aware and deadline-sensitive service characteristics. It employs a Lyapunov-based proportional-fairness resource scheduling algorithm that decomposes the long-term fairness problem into single-slot sub problems, achieving a balance between service stability and fairness. This scheduling algorithm is complemented by a block-coordinate descent method for iteratively solving non-convex fair sub problems.

\subsection{Fault-Tolerance}

Fault-tolerant scheduling strategies are essential for ensuring reliable performance especially under conditions of dynamic workload and potential failures of edge nodes.

\textbf{Checkpoint with Replication.} Xu et al.\cite{xu_cost-aware_2021} introduce a hybrid approach for low-latency stream processing that combines checkpointing with active replication of high-risk operators to balance recovery speed and resource usage. By implementing this strategy alongside RL-based dynamic scaling, the framework ensures resilient stream processing, ensuring low latency processing of IoT data streams.

\textbf{Dynamic Model Partitioning.} In contrast to stream processing focus,
FTPipeHD\cite{chen_ftpipehd_2024} extends GPU-based pipeline parallelism to edge devices for fault-tolerant DNN training. It uses dynamic model partitioning to adapt to varying device capacities and a mixed weight replication strategy for quick recovery from device failures in distributed IoT environments.

\section{Cloud-Edge Computing}

\begin{figure}[h]
    \centering
    \includegraphics[width=0.3\textwidth, page=1]{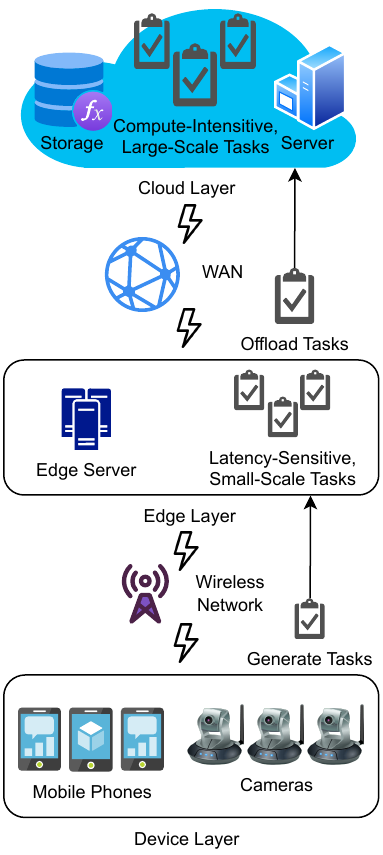}
    \caption{An architecture of the cloud \& edge collaborative computing system.}
    \label{fig:An_overview_of_the_cloud-edge_computing_system.}
\end{figure}

Cloud-edge computing combines edge processing with cloud resources, enabling tasks to be executed locally at the edge, in the cloud, or through a combination of the two. This paradigm, as shown in Fig.~\ref{fig:An_overview_of_the_cloud-edge_computing_system.}, leverages the complementary strengths of edge and cloud resources to support applications requiring both low latency and high performance. For instance, in industrial IoT, scheduling strategies focus on minimizing latency by prioritizing task execution at the edge when feasible, while offloading computationally intensive workloads to the cloud to fully utilize its capabilities. The following sections examine various scheduling methods, each designed to address specific challenges in cloud-edge computing environment.

\subsection{Performance}
\subsubsection{Computing Resources Utilization}

\textit{I. Heuristic-based Methods}

\textbf{Greedy Strategy.} 
Zhang et al.\cite{zhang_deadline-aware_2022} propose DSOTS (Dynamic Time-Sensitive Priority Algorithm) to prioritize time-sensitive tasks by analyzing submission, waiting, and execution queues. Then TSGS (Time-Sensitive Scheduling with Greedy Strategy) further optimizes by applying a greedy strategy that matches tasks to servers based on their measured processing capability, prioritizing edge servers for latency-sensitive tasks while utilizing cloud resources when edge capacity is insufficient.
For cloud-device collaborative Large Language Model (LLM) inference, 
Yang et al.\cite{yang_efficient_2024} propose a cost-latency balancing algorithm that partitions the computation load at the operator level. It starts by placing all tasks on the edge and iteratively offloads the most resource-intensive operator (e.g., linear or attention) to the cloud if the total latency exceeds the constraint, continuing until the latency constraint is met. 

\textbf{Shortest Remaining Processing Time (SRPT). } 
GeoClone\cite{wang_geoclone_2020}, a two-step online replication strategy for straggler mitigation in geo-distributed analytics. It determines both the number and placement of task replicas: it first estimates an upper bound for replicas based on available computing slots and SRPT, then selects execution sites by considering task completion progress and resource availability across geo-distributed cloud and edge servers.

\textbf{Task-Specific Model Partition.} To establish native generative AI services to enable private, timely, and personalized experiences, Yang et al.'s\cite{tian_edge-cloud_2024} method collaborates edge-cloud through task-specific model partitioning. Lightweight models are deployed to edge nodes for latency-sensitive or privacy-sensitive tasks, while complex tasks are executed in the cloud. Dynamic updates ensure model adaptation based on real-time task demands and resource availability.

\textbf{Firefly Algorithm (FA).} 
ECFA\cite{yin_ecfa_2023} is an Efficient Convergent Firefly Algorithm which improves upon the standard FA by introducing a probability-based mapping operator to convert individual fireflies into scheduling solutions, and employs a low-complexity position update strategy to enhance computational efficiency in solution exploration. ECFA provides theoretical convergence guarantees to the global best individual in the firefly space and uses parameter analysis to prevent falling into boundary traps.

\vspace{\baselineskip}
\noindent \textit{II. AI-based Methods}

\textbf{LLM-Assisted Scheduling.} For collaborative edge \& cloud LLM inferencing, Zhou et al.\cite{zhou_generative_2024} propose an in-context learning with LLMs to make offloading decisions between local and cloud processing. It uses formatted natural language descriptions, examples, and rules as prompts to guide LLMs in selecting ``local'' or ``offload'' based on service types and estimated output token sizes. To further refine the decision-making process, prioritized experience replay and epsilon-greedy strategies are applied to improve the experience pool with better examples.

\vspace{\baselineskip}
\noindent \textit{III. Mathematical Methods}

\textbf{Lyapunov-based Optimization.} Fan et al.\cite{fan_collaborative_2024} introduce a collaborative scheme for service placement, task scheduling, computing resource allocation, and transmission rate management in cloud-edge cooperative networks. It transforms the complex optimization problem into a deterministic format for each time slot using Lyapunov optimization, then employs a hybrid numerical iterative algorithm to efficiently solve it. 

\textbf{Water-Filling based.} For situation under cloud-assisted mobile edge computing, scheduling faces challenges of task arrival dynamics, edge node heterogeneity, and the trade-off between computation and communication delay. Ma et al.\cite{ma_dynamic_2023} introduce a Water-filling Based Dynamic Task Scheduling (WiDaS) algorithm leveraging the Lyapunov optimization method and a water-filling strategy to balance workloads across edge nodes and cloud. 

\vspace{\baselineskip}
\noindent \textit{IV. Hybrid Methods}

\textbf{Reinforcement Learning (RL) + Heuristic.} Kubernetes is not well-suited for deploying containers in geo-distributed computing environments and dealing with the dynamism of application workload and computing resources. 
To enable QoS-aware deployment for latency-sensitive applications, Ge-kube\cite{rossi_geo-distributed_2020} (Geo-distributed and Elastic deployment of containers in Kubernetes) extends Kubernetes through a two-step control loop: a model-based RL approach dynamically adjusts container replicas based on application response time, and a network-aware placement policy allocates containers on geo-distributed resources while considering network delays among computing resources.

\subsubsection{Data Management}
\textit{I. Data Communication Layer}

\textbf{Distributed Simulation Application.} Pond\cite{miao_efficient_2022} is a collaborative flow-based scheduler maps tasks across both cloud and edge nodes: placing computation-intensive, loosely-coupled tasks in cloud DCs while deploying user-interactive components to nearby edge nodes to reduce communication delay. By formulating this as a min-cost max-flow problem, Pond converts the task placement constraints and communication costs into network arc costs, and introduces a dominant resource method to handle multi-dimensional resource requirements.

\textbf{Real Time Streaming Analytics.}  TTL (Time To Live) is a mechanism that limits the lifespan or duration of data in a network. Kumar et al.\cite{kumar_ttl-based_nodate} propose a TTL-based data aggregation mechanism for geo-distributed streaming analytics in a hub-and-spoke edge-cloud architecture. By allocating TTL values to keys at edge servers, it optimizes the trade-off between WAN traffic and processing delay, determining how much aggregation should be performed at the edges versus the central hub. This is particularly important for applications like Akamai's media analytics, where different services require different delay-traffic balances.
\textit{Aggregation Network.} Aggnet\cite{kumar_aggnet_nodate} minimizes traffic costs through a three-tier aggregation network (edge-transit-destination) by strategically placing data aggregation operations across tiers. It formulates this as a MINLP to balance the trade-offs between traffic volume and heterogeneous regional costs, reducing cost by determining optimal aggregation points and routing paths rather than simply using nearest-neighbor routing.

\textbf{Distributed Machine Learning under Wide-Area Networks (DML-WANs).} DML-WANs faces a sequential communication dependency bottlenecks between local model computing and global model synchronization. Zhou et al.\cite{zhou_nbsync_2023} propose Non-Blocking Synchronization (NBSync) for distributed ML in edge-cloud WANs. Unlike traditional parameter server approaches that sequentially execute local computing and global synchronization, NBSync enables parallel execution of these two processes through a non-blocking synchronization mechanism. It specifically addresses the challenges of computing heterogeneity across edge servers and low WAN bandwidth between edge-cloud, achieving 1.43-2.79× speedup in training time.

\vspace{\baselineskip}
\noindent \textit{II. Data Storage Layer}

\textbf{Piggybacking.} \textit{run}Data\cite{jin_run_2022} is an online algorithm optimizes geo-distributed data analytics by coupling task offloading with data redistribution via piggybacking. It involves calculating probabilities to determine which tasks and associated data should be offloaded from edge nodes to DCs. Although \textit{run}Data may delay the execution of current jobs, it ensures hot data is efficiently relocated to reduce future overall job completion times since some datasets would be used multiple times.

\subsection{Fairness}

Fairness-oriented scheduling techniques offer mechanisms to balance performance and equitable resource allocation, ensuring proportionate access to computing resources for diverse tasks.

\textbf{Computing Offloading.} Hao et al.\cite{hao_time-continuous_2024} propose a time-continuous computing offloading algorithm that makes offloading decisions immediately upon task arrival, improving efficiency and scalability. They solve it through a DRL algorithm that decouples offloading decisions from task count - each decision only determines whether to process a single task at edge or cloud nodes. By using an \(\alpha\)-fair utility function of average task delay as the optimization objective and adjusting the MDP with past rewards, achieving effective balancing of delay and fairness in cloud-edge task scheduling.

\textbf{Fairness Knob.} Similarly, OnDisc\cite{han_ondisc_2019} introduces a fairness knob \( f \) that allows a trade-off between minimizing total weighted response time and ensures instantaneous fairness among jobs. By adjusting \( f \), OnDisc smoothly transitions from highly efficient scheduling to weighted round-robin, achieving flexible control over performance and fairness.

\subsection{Fault-Tolerance}

Fault-tolerant scheduling methods are essential for maintaining reliable operations in edge-cloud systems, ensuring task completion despite hardware failures and network instability.

\textbf{Data Replication-based Fault Tolerance.} Javed et al.\cite{javed_iotef_2020} propose Internet of Things Edge-Cloud Federation (IoTEF), a four-layer architecture that enables dynamic data processing placement between edge and cloud. Its key design uses Apache Kafka to ensure exactly-once data delivery and fault-tolerant replication across nodes, while leveraging Kubernetes federation to automatically reconfigure the processing pipeline based on available computing resources and network conditions. This unified approach allows applications to relocate computation between edge and cloud without code modifications.

\textbf{Redundant Execution-based Fault Tolerance.} Sun et al.\cite{sun_qos-aware_2020} propose Fault-Tolerance-Based QoS-Aware (FTBQA) algorithm employing two scheduling phases: primary copy placement for early task execution, and backup copy placement with minimal overlap to improve resource utilization, while an adjustment mechanism rearranges tasks after backup de-allocation to maintain system reliability.

\section{Geo-Distributed Supercomputer Computing (HPC)}
Geo-distributed supercomputer computing (HPC) paradigm coordinate tasks across globally distributed high-performance computing (HPC) nodes, which are designed specifically for tightly coupled, compute-intensive workloads. HPC system excels at solving tasks requiring massive parallelism and high-volume inter-node communication, making them ideal for applications like climate model, molecular dynamics, and Large Language Model (LLM) inference or training that involves transferring massive datasets. However, in geo-distributed settings, these systems face unique challenges, including inter-node communication delays, data transfer bottlenecks, and region-specific resource constraints. The following sections review scheduling methods tailored to HPC environments, emphasizing strategies to maximize efficiency and scalability. 

\subsection{Performance}

\subsubsection{Computing Resources Utilization}

\textit{I. Heuristic-based Methods}

\textbf{Ant Colony Optimization (ACO).} Cross-region inter-connection super-computing (CIS) \cite{han_accelerate_2022} is a framework that integrates geo-distributed super-computing and storage resources to meet increasing demands of tasks. The scheduling problem is modeled with constraints on deadlines and storage. They use ACO's parallel independent search method to shorten search time and improve reliability in finding optimal solutions, achieving 12.9\% shorter completion time compared to FCFS and Min-Min algorithms.

\textbf{Scheduled Neighbors Lookup (SNL).} In-situ workflows enable concurrent execution of components with continuous data flow, where performance is limited by the slowest component or data transfer. For scheduling such workflows in geo-distributed HPC environments, Li et al.\cite{li_efficient_2023} propose the SNL algorithm that creates a blended cost-sorted list of computation and communication pairs, optimizes deployment through scheduled-neighbors location analysis, and uses a refinement stage to adjust resource allocation for maximizing workflow throughput.

\vspace{\baselineskip}
\noindent \textit{II. AI-based Methods}

\textbf{Reinforcement Learning (RL).} Recent studies in optimizing energy consumption are inherently hardware-based or require profiling information in advance. Mamun et al.\cite{mamun_intra-_2020} propose a RL approach that differs from traditional hardware-based solutions like VM consolidation and Dynamic Voltage and Frequency Scaling (DVFS). Their approach dynamically schedules tasks without requiring prior job profiling information, using a Multi-Armed Bandit (MAB) model to explore and exploit job allocation patterns, while optimizing both profit through value-based scheduling and energy through intelligent resource allocation.
\textit{Direct Future Prediction (DFP)}, an Intel-developed algorithm that extends RL with dynamic goal adjustment capability, has shown success in gaming domains but remains unexplored in scheduling under HPC environment.
Li et al.\cite{li_mrsch_2022} propose an intelligent scheduling agent named \textit{MRSch} for multi-resource scheduling leveraging DFP. MRSch replaces traditional image-based encoding with a vector-based mechanism to handle HPC's widely varying job durations, while dynamically adjusting resource weights based on demand patterns. They incorporate a window-based reservation technique that combines back-filling with resource reservation, effectively preventing large job starvation while ensuring high resource utilization.

\vspace{\baselineskip}
\noindent \textit{III. Mathematical Methods}

\textbf{Annihilating Polynomial-based.} ExaLB\cite{mollasalehi_exalb_2023} is a mathematical framework for load balancing that uses annihilating polynomials to classify and schedule tasks in Distributed Exascale Computing Systems (DECS) based on dual-event types (formal vs. dynamic/interactive). Through polynomial transformations between process requests and resource capabilities, it dynamically maps tasks to resources without requiring predetermined scheduling patterns and enables adaptive load balancing in cross-domain HPC environments.

\textbf{Mixed Integer Programming.} Arabas et al.\cite{arabas_modeling_2021} propose a hierarchical task allocation framework that formulates the geo-distributed HPC scheduling as a mixed integer programming centralized problem, decomposing it into parallel sub-problems for local clusters. The framework converts the global energy minimization into binary decision variables for task allocation and power states, enabling distributed optimization through CPLEX solver while considering both computational resources and network traffic constraints.

\vspace{\baselineskip}
\noindent \textit{IV. Hybrid Methods}

\textbf{Deep Reinforcement Learning + Greedy Approach.} Yang et al.\cite{yang_deep_2021} propose a two-stage scheduling algorithm enhanced by deep reinforcement learning (DRL) for task sequencing and greedy optimization for task allocation in cloud-based HPC systems. The DRL module predicts the optimal task allocation sequence for each batch, while the greedy strategy allocates tasks online to maximize system gain with a proven competitive ratio.

\textbf{Multi-heuristic.} Traditional algorithms like Cuckoo search (CS) may be stuck in local minima, lack solution diversity and suffer from slow convergence. Chhabra et al.\cite{chhabra_performance-aware_2021} propose a multi-objective hybrid scheduling algorithm (MOHCSFA) to overcome these limitations of the traditional algorithms. It combines the solution search mechanisms of both CS and firefly algorithm during generation and further integrated with efficient resource allocation heuristic to improve scheduler performance. Similarly, Chhabra et al.\cite{chhabra_multi-criteria_2021} propose another strategy, CSDEO, which combines CS, differential evolution and Opposition-Based Learning (OBL) method to improve overall makespan and energy consumption. It first uses OBL to produce an initial population and then switches between CS exploration phase and DE exploration phase based on each solution's fitness.

\subsubsection{Data Management}

When performing high-performance computing in a wide area network (WAN) environment, the data transmission problem in distributed computing is increasingly prominent because of the geographic dispersion of super-computing centers, the complexity of the interconnection network topology, and the need to transmit a large amount of data while the WAN bandwidth is not sufficient.

\textbf{I/O Proxy.} Suffering from performance bottlenecks in data migration and access across the WAN, Huo et al.\cite{huo_research_2024} propose a multi-task-oriented data migration (MODM) method to select the appropriate data source and dynamically adjust bandwidth allocation among all migration tasks, and the request access-aware I/O proxy resource allocation (RAAS) strategy to allocate I/O proxy and optimize delay.

\subsection{Fairness}

Fairness-focused scheduling reallocates resources based on defined criteria, ensuring equitable access across competing tasks.

\textbf{Priority-based.} Posner et al.\cite{posner_enhancing_2023} propose a malleable job scheduling strategy for supercomputers, centered on fairness in resource allocation. The approach defines three priority criteria—based on job age, remaining runtime, and resource usage history—to decide which malleable jobs should receive resource reassignment first. Additionally, it introduces three strategies for timing resource reassignments: immediate, delayed, and gradual, which manage the interval and smoothness of resource transfer between jobs.

\subsection{Fault-Tolerance}

Fault-tolerance strategies in HPC systems increasingly incorporate energy-efficient methods to balance reliability with reduced energy consumption during recovery from failures.

\textbf{Rollback-Recovery.} Morán et al.\cite{moran_towards_2020} introduce a set of strategies aiming to enhance energy efficiency in fault-tolerant HPC systems by focusing on reducing energy consumption during failures using rollback-recovery methods with uncoordinated checkpoints. The strategies target nodes that do not need rollback and explore the use of Dynamic Voltage and Frequency Scaling (DVFS) and system hibernation techniques. A specially designed simulator, now extended with non-blocking communication capabilities\cite{moran_exploring_2024} and an increased number of candidate processes for analysis, evaluates these strategies to identify the most effective energy-saving approaches.

\section{Challenges and Open Issues}
Aiming to assist researchers interested in geo-distributed computing and to promote deeper investigation into this domain, this section explores the research challenges, potential opportunities, and unresolved issues related to task scheduling in geo-distributed computing systems.

\textbf{Emerging Workload Diversity.} Due to the increasing diversity in application types, geo-distributed computing faces growing scheduling challenges. Most existing scheduling approaches, while effective for traditional high-performance computing (HPC) and web service applications, struggle to handle emerging workload types, such as AI, big data and multi-modal computational paradigms. These emerging workloads necessitate innovative scheduling strategies tailored for geo-distributed computing environments and capable of effectively exploiting the distributed computational capacities of geo-distributed infrastructures.

For instance, applications such as LLM inference and AR/VR-integrated intelligent assistants (e.g., ChatGPT’s video-calling mode\cite{openai_santamode}) require coordination of multi-modal tasks, cross-region computational coordination, and the ability to leverage heterogeneous hardware such as CPUs, GPUs, and specialized accelerators. Similarly, time-sensitive AI applications, including conversational AI services (e.g., ChatGPT’s voice-calling mode, 1-800-CHATGPT hotline)\cite{openai_1800chatgpt}, require real-time response from servers. Both types of applications need strict adherence to Quality of Service (QoS) metrics, but often suffer from capacity limitations. 

These challenges are amplified in geo-distributed environments, where resource-demand imbalances, multi-stage processing pipelines, and network dynamics introduce additional complexity to workload scheduling. Existing scheduling approaches still remain insufficient to optimally allocate resources across geo-distributed regions, effectively manage intricate task dependencies, and dynamically adapt to real-time application requirements.

\textbf{Next Generation Geo-Distributed Computing.} As computational hardware continues to advance, the next generation of geo-distributed computing is prepared to incorporate nontraditional hardware architectures, such as quantum computing and nano-computing. These cutting-edge technologies promise to revolutionize computational power and efficiency, offering unprecedented capabilities. 

Nano-computing focuses on developing computational devices at the molecular and atomic scales. It enables unprecedented miniaturization through innovative materials and architectural designs, dramatically reducing physical footprint and energy consumption. Quantum computing leverages quantum mechanical phenomena like superposition and quantum entanglement to perform parallel computations. This approach enables solving complex optimization and cryptographic problems that are computationally infeasible for classical systems. Additionally, quantum communication, a critical aspect of quantum computing technology, leverages quantum entanglement and quantum key distribution. This mechanism enables both ultra-secure and ultra-fast data transmission, redefining how information is shared across distributed computing systems. These are representative technologies shaping the future of high-performance geo-distributed computing environments. 

However, employing these novel hardware architectures and integrating these advanced technologies into existing computing infrastructures poses significant challenges, including the development of specialized hardware architectures, software frameworks, resource management methodologies and task scheduling strategies. This transition will require significant innovation to fully realize the potential of these emerging technologies.

\textbf{Security and Privacy.} Geo-distributed tasks often involve handling massive amounts of data generated from multiple geo-distributed locations. Ensuring secure data transmission and compliant task execution has become a critical issue. These challenges are exacerbated in geo-distributed computing environments, particularly when managing sensitive data across multiple jurisdictions and heterogeneous edge devices. The distributed nature of these systems introduces vulnerabilities at various levels, including data transmission between nodes and computation on untrusted edge devices. 

Existing security mechanisms are limited in addressing these challenges due to the resource constraints of edge devices, the complexity of enforcing consistent security policies across diverse geographical regions with varying regulatory requirements, and the overhead of cryptographic operations in real-time applications. This necessitates the development of new security-aware scheduling algorithms and related mechanisms that incorporate regional compliance requirements and ensure secure data handling during task distribution and execution.

\section{Conclusion}

Task scheduling in geo-distributed computing has attracted significant attention from both industry and academia due to its potential to leverage global distributed computational resources and execute large-scale computational tasks. However, most existing surveys on task scheduling fail to differentiate between specific geo-distributed computing infrastructures. To address this gap, we present a comprehensive review of state-of-the-art task scheduling techniques across four distinct geo-distributed computing systems. We categorize scheduling algorithms based on different scheduling objectives (performance, fairness, fault-tolerance). Finally, we discuss the key challenges and open research issues in this field. We aim for this survey to serve as a valuable resource for researchers and practitioners, guiding continued exploration and innovation in this domain.


%

\ifCLASSOPTIONcaptionsoff
  \newpage
\fi



%

\bibliography{References}
\bibliographystyle{IEEEtran}


%




\end{document}